\documentclass[12pt]{article}
\usepackage{a4wide}
\usepackage{graphicx}
\usepackage{amssymb}
\usepackage{amsmath}
\usepackage{slashed}
\usepackage{cite}
\usepackage{multicol}
\usepackage{braket}
\usepackage{bmpsize}
\usepackage{xcolor}
\usepackage{subcaption}
\usepackage{tablefootnote}
\usepackage{float}

\newcommand{\be}{\begin{equation}}
\newcommand{\ee}{\end{equation}}
\newcommand{\ba}{\begin{eqnarray}}
\newcommand{\ea}{\end{eqnarray}}

\begin{document}

\allowdisplaybreaks

\begin{titlepage}
\begin{flushright}
\end{flushright}
\vfill
\begin{center}
{\Large\bf Multi Component Dark Matter in a Minimal Model}
\vfill
{\bf Karim Ghorbani}\\[.5cm]
{{\it Physics Department, Faculty of Science, Arak University, Arak 38156-8-8349, Iran}}\\[1cm]

\end{center}
\vfill
\begin{abstract}
We study a minimal $\mathbb{Z}_2$-symmetric extension of the Standard Model containing two singlet fermions and a singlet scalar that interact with the SM particles through the Higgs-portal. We identify regions of parameter space in which all three new particles are kinematically stable, giving rise to a multi component dark matter (DM) scenario.
The parameter space consistent with the observed dark matter relic abundance are determined, and the contribution of each component to the total relic density is evaluated. While the DM–nucleon elastic scattering cross sections of the two fermionic dark matter components are loop-suppressed, the corresponding cross section of the scalar dark matter particle arises at tree level and is therefore expected to dominate.
We find a viable region of the parameter space in which the scalar 
dark matter candidate with mass range of approximately $125-400$ GeV, evades current direct detection (DD) bounds while contributing only a small fraction of the observed relic density. In contrast, the fermionic dark matter possesses a loop-suppressed DD cross section that lies below the neutrino floor and can constitute a substantial fraction of the total relic density.

\end{abstract}
\vfill

\vfill
{\footnotesize\noindent }

\end{titlepage}

\section{Introduction}

A wide range of astrophysical and cosmological observations provide compelling evidence for the existence of dark matter, which constitutes about $27\%$ of the energy density of the Universe \cite{Bergstrom:2000pn,Feng:2010gw}. 
Despite its well-established gravitational imprint, the particle nature of dark matter remains unknown, motivating extensions of the Standard Model (SM) that can account for its stability, relic abundance, and interactions with visible matter. Weakly interacting massive particles (WIMPs) are well-known dark matter candidates \cite{Arcadi:2017kky}. A widely studied mechanism for dark matter production is the freeze-out paradigm, in which dark matter particles are assumed to be in thermal equilibrium with 
the SM particles in the early Universe before decoupling 
as the Universe expands and cools \cite{Steigman:2012nb}.

Among the simplest extensions of the Standard Model are Higgs-portal models, in which the dark sector communicates with the SM particles through operators involving the Higgs doublet. 
Such models are attractive due to their minimal field content and renormalizability, while remaining testable through collider searches, invisible Higgs decays, and direct and indirect detection experiments. Nevertheless, many Higgs-portal scenarios are increasingly constrained by direct detection experiments, especially when dark matter interaction with the SM particles is mediated by the Higgs boson at tree level \cite{McDonald:1993ex,Kim:2008pp}.
An alternative approach is provided by pseudoscalar mediators, which naturally suppress spin-independent direct detection cross sections and are therefore less constrained by current experimental bounds. Such models have been widely explored in the context of fermionic dark matter \cite{Ghorbani:2014qpa,Fan:2015sza,Yang:2016wrl,Baek:2017vzd,Ghorbani:2017jls,Ghorbani:2018pjh,DiazSaez:2021pmg,Chen:2024njd}.


In this work, we study a minimal extension of the SM consisting of two gauge-singlet Dirac fermions and a real scalar mediator. The model is endowed with an exact $\mathbb{Z}_2$ symmetry under which one of the fermions and the scalar field are odd, while the second singlet fermion and all SM fields are even. 
As a consequence of this symmetry, fermion mixing and linear coupling between the scalar field and the Higgs sector are forbidden, ensuring a vanishing vacuum expectation value for the singlet scalar and the absence of scalar-Higgs mixing.
The only renormalizable interaction connecting the dark sector to the SM arises through a Higgs-portal operator involving pairs of scalar fields.
The exact $\mathbb{Z}_2$ symmetry therefore leads to a minimal framework in terms of both field content and the number of free parameters.

In the model described above, two kinematically distinct dark matter configurations are possible: a two-component and a three-component dark matter scenario. 
In the former case, the dark sector consists either of one fermion and one scalar or of two fermions.
In \cite{Bhattacharya:2013hva} they studied a related two-component scenario involving a Majorana fermion and a singlet scalar, where the heavier fermion is unstable. 
Their analysis showed that, apart from a resonance region, the scalar dark matter candidate is excluded by current direct detection constraints.
We confirm this conclusion even after including all relevant interactions in the relic density computation.
The second possibility arises when all three singlet states are kinematically stable, thereby forming a three-component dark matter sector. This is the scenario considered in the present work. The resulting phenomenology is governed by dark sector interactions and Higgs-portal processes. In particular, the direct detection cross sections of the fermionic dark matter components are loop-suppressed and remain below the neutrino floor, whereas the scalar dark matter candidate can possess a direct detection cross section lying below the current experimental bounds while remaining within reach of future searches. Consequently, this framework provides a simple realization of a scenario in which the dominant dark matter component may evade direct detection, while a subdominant scalar component remains experimentally accessible.

The remainder of this paper is organized as follows. In Sec.~\ref{model}, we present the details of the model and discuss its symmetry structure. In Sec.~\ref{3DMScenario}, we discuss the three-component dark matter scenario and introduce the interactions relevant for the relic density calculation. The corresponding Boltzmann equations are presented, and viable regions of the parameter space consistent with the observed relic abundance are identified. In Sec.~\ref{DDCS}, we calculate the DM–proton elastic scattering cross sections. After imposing the relic density constraint from the Planck measurement, we find regions of the parameter space consistent with the upper bounds from XENON1T and XENONnT. Finally, we summarize our results and conclude in Sec.~\ref{Conc}.

\section{Model}
\label{model}
We consider an extension of the Standard Model by incorporating two gauge-singlet Dirac fermions, $\psi_1$ and $\psi_2$, and a real scalar field $\phi$. The model is endowed with an exact discrete $\mathbb{Z}_2$ symmetry under which the fields transform as
\begin{equation}
\psi_1 \;\rightarrow\; -\psi_1, 
\qquad 
\psi_2 \;\rightarrow\; \psi_2, 
\qquad 
\phi \;\rightarrow\; -\phi,
\end{equation}
whereas all the SM fields, including the Higgs doublet $H$, are taken to be $\mathbb{Z}_2$-even.

The most general renormalizable Lagrangian consistent with the gauge symmetries of the SM and the imposed $\mathbb{Z}_2$ symmetry contains three interaction terms
\begin{equation}
\mathcal{L} \;\supset\;
y\,\bar{\psi_1}\,\psi_2\,\phi
\;-\;
\lambda\,\phi^2\,H^\dagger H
\;-\; \kappa \, \phi^4
\;+\;
\text{h.c.},
\end{equation}
together with the canonical kinetic and mass terms,
\begin{equation}
\mathcal{L} = 
\sum_{i=1,2 }\bar{\psi}_i\left(i\gamma^{\mu}\partial_{\mu}-m_{i}\right)\psi_i + \frac{1}{2} \partial^{\mu} \phi \partial_{\mu} \phi - \frac{1}{2} m_\phi^2 \phi^2.
\end{equation}
The discrete symmetry forbids linear and cubic terms in $\phi$, as well as fermion mixing term $\bar{\psi_1}\psi_2$, ensuring that the vacuum expectation value of $\phi$ vanishes and that there is no mixing between 
$\phi$ and the SM Higgs boson. Consequently, the Higgs-portal operator $\phi^2 H^\dagger H$ provides the only renormalizable interaction between the dark sector and the SM.
After electroweak symmetry breaking, one can expand the Higgs field as
\begin{equation}
H = \frac{1}{\sqrt{2}}
\begin{pmatrix}
0 \\ v_h + h
\end{pmatrix},
\end{equation}
where $v_h = 246$ GeV is the Higgs vacuum expectation value. 
The interaction term, $\phi^2 H^\dagger H$, generates an additional contribution to the scalar mass. Consequently, the physical mass of $\phi$ becomes $m^2_{\text{phy}} = m_\phi^2 + \frac{1}{2}\lambda v_h^2$.
The Higgs-portal interaction induces couplings between pairs of $\phi$ fields and the physical Higgs boson while preserving the $\mathbb{Z}_2$ symmetry.

\section{Three Dark Matter Scenario}
\label{3DMScenario}
\begin{figure}[H]
\centering
\includegraphics[width=.8\textwidth,angle =0]{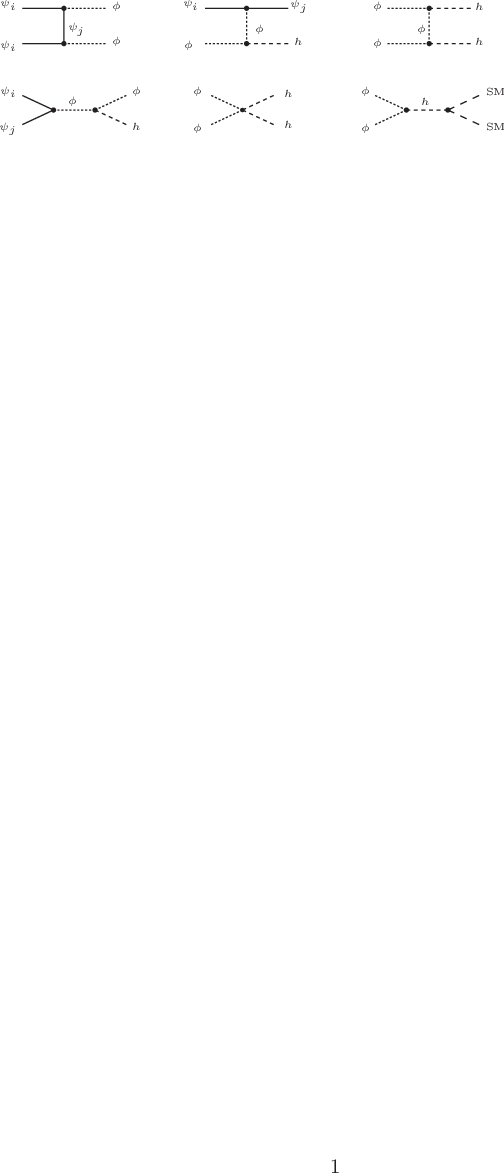}
\caption{Relevant Feynman diagrams contributing to the relic density calculation.\protect\footnotemark}
\label{relic_diagram}
\end{figure}

We focus on the region of parameter space in which the three new particles, namely the singlet Dirac fermions $\psi_1$, $\psi_2$ and the singlet scalar $\phi$, are kinematically stable and can therefore constitute dark matter candidates.
To this end, the following conditions are required simultaneously, 
\begin{equation}
m_1 < m_\phi + m_2, ~~ m_2 < m_\phi + m_1,~~ m_\phi < m_1 + m_2 \,.
\end{equation}
We now discuss the interactions relevant for the relic density calculation.
Moreover, the relevant Feynman diagrams are depicted in Fig.~\ref{relic_diagram}.


\subsection{Scalar DM Processes}
We provide number changing processes for the scalar DM which are 
kinematically allowed. 
The scalar dark matter candidate participates in both self-annihilation and conversion processes.
Self-annihilation proceeds through the channels $\phi\phi\to\psi_i\psi_i$ via $t$- and $u$-channel by the 
exchange of $\psi_j$. In addition, the scalar can annihilate into the SM particles through Higgs and scalar mediated interactions. The process $\phi\phi\to{\rm SM}$ receives contributions from the $s$-channel with Higgs exchange and from the $t$- and $u$-channel with scalar exchange. A contact interaction also contributes to the process $\phi\phi\to hh$.
Conversion processes include reactions such as $\phi\psi_i\to h\psi_j$ and $\phi h\to\psi_1\psi_2$.

\subsection{Fermion DM Processes}
The fermionic dark matter components participate in self-annihilation, conversion, and 
co-annihilation processes whenever these channels are kinematically accessible.
Self-annihilation proceeds through the interaction $\psi_i\psi_i\to\phi\phi$ via $t$- and $u$-channel by the exchange of $\psi_j$. Conversion reactions include $\psi_i\phi\to\psi_j h$ and $\psi_i h\to\psi_j\phi$, mediated by scalar exchange. 
In addition, a co-annihilation process as $\psi_i\psi_j\to\phi h$ can occur through the $s$-channel
by exchanging a singlet scalar.
\footnotetext{The package {\tt Axodraw} is used to create the Feynman diagrams \cite{Collins:2016aya}.}

\subsection{Boltzmann Equations}
In this model, the three DM components are $\psi_1, \psi_2$, and $\phi$, 
with number densities $n_1, n_2$, and $n_\phi$, respectively.
The evolution of the dark matter number densities is governed by a coupled set of 
Boltzmann equations, which can be written as (see conventions in \cite{Belanger:2014vza}), 
\begin{align}
\frac{dn_\phi}{dt} + 3H n_\phi
&=
- \langle \sigma v \rangle_{\phi \phi \to \text{SM}}
\left(
n_\phi^2 - \bar n_\phi^2 \right)
- \langle \sigma v \rangle_{\phi \phi \to 11}
\left(
n_\phi^2 - n_1^2 \frac{\bar n_\phi^2}{\bar n_1^2}
\right)
\nonumber \\
&\quad
- \langle \sigma v \rangle_{\phi \phi \to 22}
\left(
n_\phi^2 - n_2^2 \frac{\bar n_\phi^2}{\bar n_2^2}
\right)
- \frac{1}{2} \langle \sigma v \rangle_{\phi 1 \to  h2}
\left(
n_\phi n_1 - n_2 \frac{\bar n_\phi \bar n_1}{\bar n_2}
\right)
\nonumber \\
&\quad
- \frac{1}{2} \langle \sigma v \rangle_{\phi 2 \to  h1}
\left(
n_\phi n_2 - n_1 \frac{\bar n_\phi \bar n_2}{\bar n_1}
\right)
- \frac{1}{2} \langle \sigma v \rangle_{\phi h \to  12}
\left(
n_\phi - n_1 n_2 \frac{\bar n_\phi}{\bar n_1 \bar n_2}
\right),
\\[6pt]
\frac{dn_1}{dt} + 3H n_1
&=
- \langle \sigma v \rangle_{11\to \phi \phi}
\left(
n_1^2 - n_\phi^2 \frac{\bar n_1^2}{\bar n_\phi^2}
\right)
- \frac{1}{2}  \langle \sigma v \rangle_{\phi 1 \to h2}
\left(
n_1 n_\phi - n_2 \frac{\bar n_1 \bar n_\phi}{\bar n_2}
\right) 
\nonumber \\
&\quad
- \frac{1}{2}  \langle \sigma v \rangle_{h 1 \to \phi 2}
\left(
n_1 - n_2 n_\phi \frac{\bar n_1}{\bar n_2 \bar n_\phi}
\right)
- \frac{1}{2} \langle \sigma v \rangle_{12 \to \phi h}
\left(
n_1 n_2 - n_\phi \frac{\bar n_1 \bar n_2 }{\bar n_\phi}
\right),
\\[6pt]
\frac{dn_2}{dt} + 3H n_2
&=
- \langle \sigma v \rangle_{22\to \phi \phi}
\left(
n_2^2 - n_\phi^2 \frac{\bar n_2^2}{\bar n_\phi^2}
\right)
- \frac{1}{2} \langle \sigma v \rangle_{\phi 2 \to h 1}
\left(
n_2 n_\phi - n_1 \frac{\bar n_2 \bar n_\phi}{\bar n_1}
\right)
\nonumber \\
&\quad
- \frac{1}{2}  \langle \sigma v \rangle_{h 2 \to \phi 1}
\left(
n_2 - n_1 n_\phi \frac{\bar n_2}{\bar n_1 \bar n_\phi}
\right)
-  \frac{1}{2} \langle \sigma v \rangle_{12 \to \phi h}
\left(
n_1 n_2 - n_\phi \frac{\bar n_1 \bar n_2 }{\bar n_\phi}
\right),
\end{align}
where $H$ is the Hubble parameter and the quantity 
$\langle \sigma v \rangle_{ij\to kl}$ denotes the thermally averaged 
cross section for the process $ij\to kl$.
The quantity $\bar n_\alpha$ denotes the equilibrium number density of the non-relativistic 
species $\psi_{\alpha}$ or $\phi$.  
We assume that the dark matter particles scatter off the SM particles 
frequently enough such that their temperature coincides with that of the Standard Model plasma.
The relic abundance is therefore determined by the interplay of self-annihilation, 
conversion, and co-annihilation processes, leading to freeze-out dynamics that differ 
qualitatively from those of conventional WIMP scenarios dominated solely by self-annihilation.

\begin{figure}
\centering
\includegraphics[width=.5\textwidth,angle =-90]{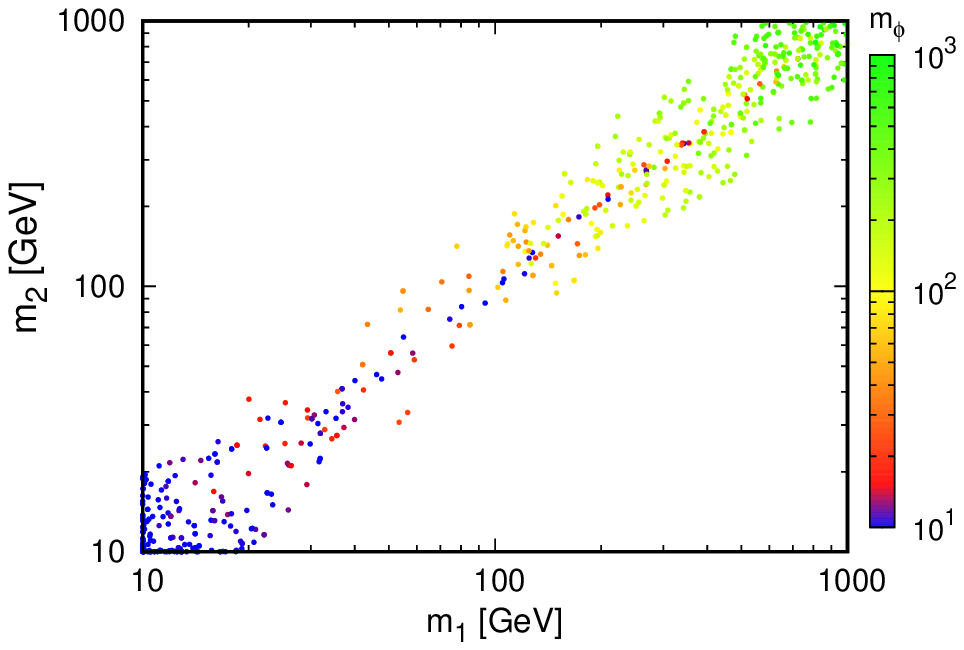}
\caption{Viable regions of the mass parameter space consistent with the observed dark matter relic abundance.} 
\label{3DM}
\end{figure}
\begin{figure}
\centering
\includegraphics[width=.5\textwidth,angle =-90]{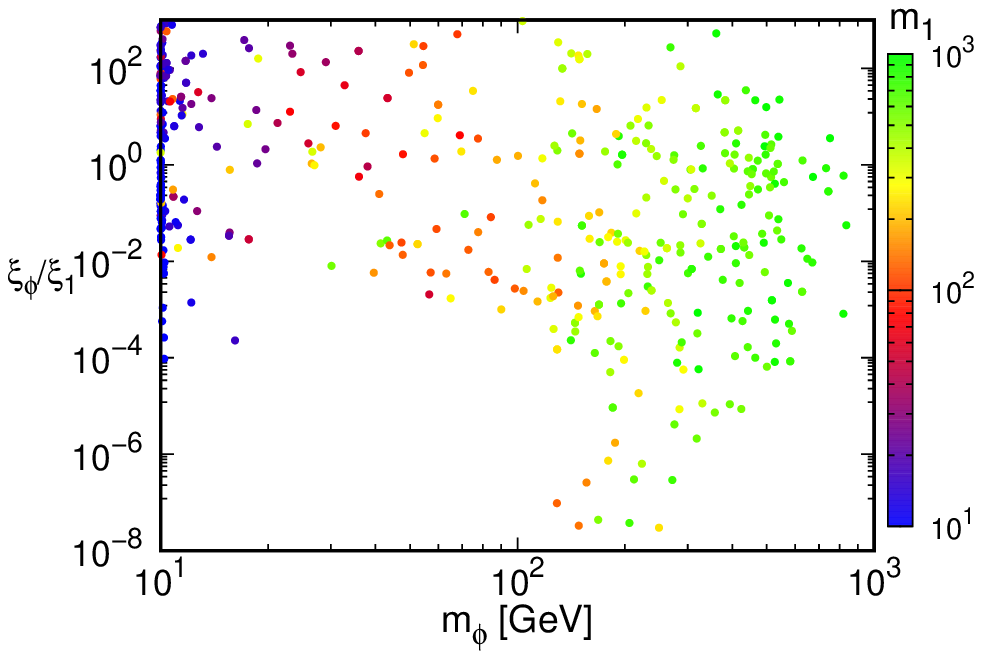}
\caption{The ratio $\xi_\phi/\xi_1$ as a function of the scalar dark matter mass. The color scale indicates the corresponding fermion dark matter mass.} 
\label{RelativeFraction}
\end{figure}

\subsection{Numerical Analysis for Relic Density}
In this subsection, we perform numerical analysis of the dark matter relic abundance in the 
model parameter space. 
To this end, we employ the package {\tt micrOMEGAs} \cite{Alguero:2023zol}, in which these processes 
are taken into account in order to compute 
the relic density: self-annihilation, conversion, and co-annihilation. 
As input parameters we have three DM masses, $m_1, m_2, m_\phi$, and two couplings $y$ 
and $\lambda$. 
We perform a parameter-space scan to identify points respecting the observed relic density, $\Omega_{\text{DM}} h^2 = 0.120 \pm 0.001$ \cite{Planck:2018vyg}.
The parameter ranges used in the scan are, $10~\text{GeV} < m_i < 1~\text{TeV}$ and $0 < y, \lambda < 2$.
In our numerical analysis we set $\kappa = 0$, since this interaction does not contribute to the relic density or direct detection observables. Furthermore, because we consider only regions with $\lambda > 0$, the bounded-from-below condition is satisfied.
The total relic abundance receives contributions from all three dark matter 
components, with each component contributing a fraction $\xi_i=\Omega_i/\Omega_{\rm DM}$.
As a first step, we find viable dark matter masses by imposing the constraint from the observed relic density on the predicted relic abundance. 
Fig.~\ref{3DM} shows a clear correlation between the fermion and scalar masses, with 
larger fermion masses generally favoring larger scalar masses in order to reproduce 
the observed relic abundance.
We next examine the ratio $\xi_\phi/\xi_1$ as a function of the scalar mass
and the mass of a fermion. 
The results shown in Fig.~\ref{RelativeFraction} divide the parameter 
space into regions with $\xi_\phi>\xi_1$ and $\xi_\phi<\xi_1$.
In particular, for $m_\phi > 125$ GeV, the ratio $\xi_\phi/\xi_1$ drops below $10^{-4}$. 
This behavior arises because the annihilation channel $\phi\phi\to hh$ becomes kinematically 
accessible once the scalar mass exceeds the Higgs mass. 
The resulting enhancement of the annihilation rate reduces the relic abundance of 
the scalar dark matter component, consistent with the approximate relation $\Omega_i \propto 1/\langle\sigma_i v\rangle$.

\section{Direct Detection Cross Section}
\label{DDCS}
\begin{figure}
\hspace{2.1cm}
\begin{minipage}{.5\textwidth}
\includegraphics[width=.6\textwidth,angle =0]{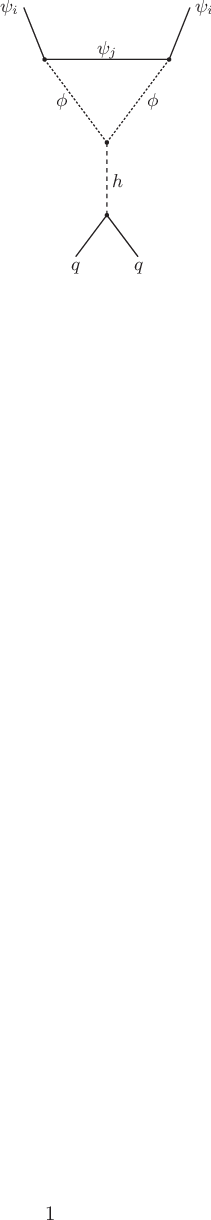}
\end{minipage}
\hspace{.1cm}
\begin{minipage}{.4\textwidth}
\includegraphics[width=.4\textwidth,angle =0]{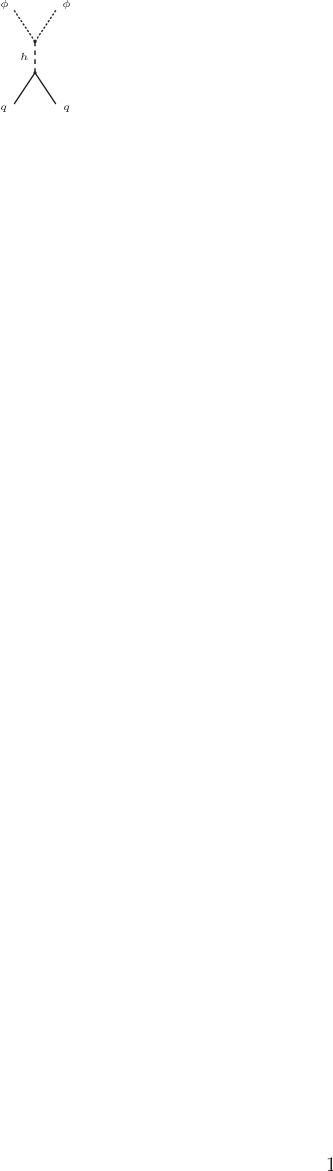}
\end{minipage}
\caption{Feynman diagrams contributing to DM–quark elastic scattering. 
The left panel shows the loop-induced interaction for fermionic 
dark matter, while the right panel depicts the tree-level Higgs-mediated 
interaction for scalar dark matter.} 
\label{DD_diagrams}
\end{figure}

In this section, we provide the DM-proton elastic scattering cross section for the fermionic and scalar dark matter components. The relevant Feynman diagrams for the scalar and fermion dark matter are depicted in Fig.~\ref{DD_diagrams}
\subsection{Fermion DM}
The elastic scattering of fermionic dark matter off nucleons is absent at tree level.
For a fermionic dark matter particle $\psi_i$ with mass $m_i$, the DM–quark scattering 
amplitude is generated at one loop through a triangle diagram involving two scalar 
fields $\phi$ and a fermion $\psi_j$. The scattering amplitude reads,
\begin{equation}
  i{\cal M} = 
  \Big[\frac{C_h}{(p_1-p_2)^2-m_{h}^2} \Big] \bar q q~  \times 
\int \frac{d^4q}{(2\pi)^4}  \frac{  y^2 C_{\phi \phi h}~~\bar \psi_i(p_{2}) (\slashed{q}+m_{j}) \psi_i(p_1) }
{[(p_2-q)^2-m_{\phi}^2][(p_1-q)^2-m_{\phi}^2][q^2-m_{j}^2]}   \,.
\end{equation}
In the expression above, 
we have $C_h= -m_q/v_h$, and $C_{\phi \phi h} = 2 \lambda v_h$ is the trilinear scalar coupling.
The corresponding effective scattering amplitude in the limit that the 
momentum transferred to a nucleon is $q^2 \sim 0$, can be written as 
\begin{equation}
 {\cal M}_{\text{eff}} =  \frac{m_{j} y^2 C_{\phi \phi h}}{16\pi^2} 
 (\frac{m_q}{v_h}) \frac{1}{m_h^2} {\cal F(\beta)} ~(\bar q q)(\bar \psi_i \psi_i) 
   \equiv m_q ~\alpha ~(\bar q q)(\bar \psi_i \psi_i)
 \,,
\end{equation}
in which $\beta = m_\phi^2/m_{j}^2$ and the loop function ${\cal F(\beta)}$ 
is given below,
\begin{equation}
\begin{aligned}
{\cal F(\beta)} {} & =  -\frac{1}{4m_{j}^2} + \frac{1}{4(\beta-4)m_{j}^2} \Big[
            2(\beta-3)\sqrt{\beta^2-4\beta} \log \frac{\sqrt{\beta-4}+\sqrt{\beta}}{\sqrt{\beta-4}-\sqrt{\beta}} 
\\ & 
 + 2(\beta^2-5\beta+4) \log \beta 
  +2(\beta-3) \log \frac{\beta^2-4\beta-(\beta-2)\sqrt{\beta^2-4\beta}}{\beta^2-4\beta+(\beta-2)\sqrt{\beta^2-4\beta}} \Big] 
  \,.      
\end{aligned}
\end{equation}
An analogous loop function was derived in  \cite{Ghorbani:2018pjh}.
The resulting spin-independent DM–proton scattering cross section is
\begin{equation}
\sigma^p_{\text{fermion}} = \frac{4\alpha_p^2\mu^2}{\pi} \,,
\end{equation}
where $\mu$ is the reduced mass of the proton and the DM, and 
$\alpha_p$ is 
\begin{equation}
 \alpha_p = m_{p} \alpha \Big( \sum_{q = u,d,s} f^{p}_{q}  
+ \frac{2}{9} f^{p}_{g}   \Big)  \equiv m_{p} \alpha f_{p} \,,
\end{equation}
where $m_p$ is the proton mass, while $f_q^p$ and $f_g^p$ denote the nucleon 
scalar form factors.
 Numerically, $f_{p} \sim 0.285$ \cite{Alguero:2022inz}.

\subsection{Scalar DM}
The elastic scattering of scalar dark matter off nucleons occurs at tree level through Higgs exchange.
The resulting spin-independent cross section is given by
\begin{equation}
 \sigma^p_{\text{scalar}} = \frac{m^4_{p}\lambda^2 f^2_p}{\pi m^4_{h}(m_p + m_\phi)^2} \,.
\end{equation}

\subsection{Numerical Results for Direct Detection}
\begin{figure}
\centering
\includegraphics[width=.49\textwidth,angle =-90]{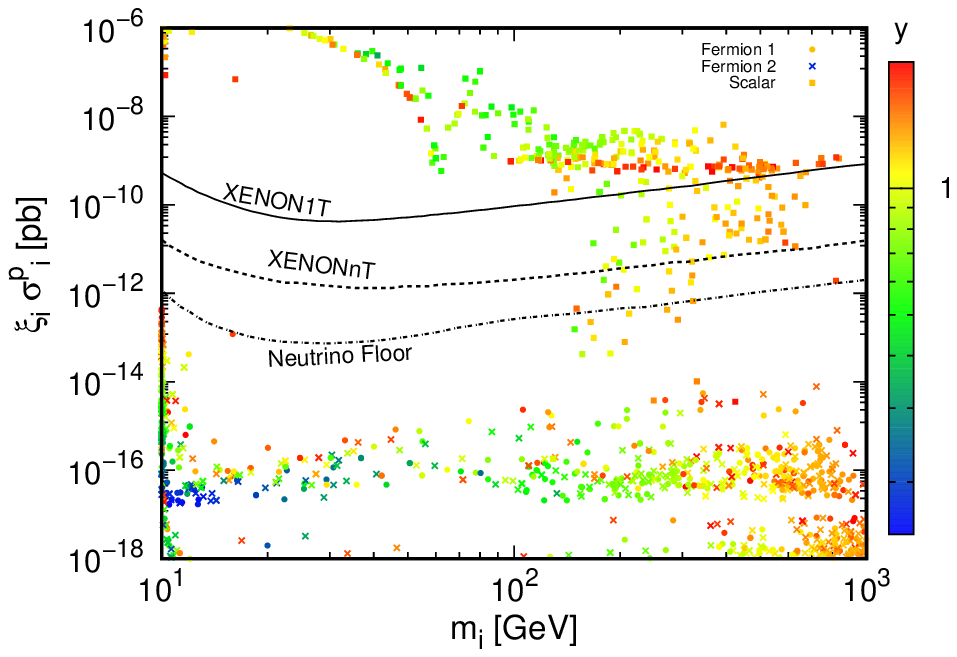}
\caption{The DM–nucleon elastic scattering cross sections of the three dark matter components are shown. 
The cross sections are rescaled by the fraction of relic density of the respective DM components. Upper bounds from XENON1T and XENONnT, and lower bound from neutrino floor are placed.} 
\label{DD_results}
\end{figure}

We now present the numerical results for the direct detection cross sections of the 
three dark matter components.
The range of the parameters we apply in our scan is the same as before. 
As expected, over the entire scalar mass range, $\sigma^{p}_{\text{scalar}} \gg \sigma^{p}_{\text{fermion}}$. 
This is because the cross section for fermions is loop suppressed. 
The direct detection cross section of each component must be rescaled by its relic density fraction,
$\xi_{i} \times \sigma^{p}_{i}$, and then confront it against the upper bounds from direct detection experiments, XENON1T \cite{XENON:2018voc} and XENONnT \cite{XENON:2023cxc}. 
Our results for the DD cross section of each DM component are presented in Fig.~\ref{DD_results}.
Moreover, the expectation based on the results obtained from the singlet scalar model is that the entire range of the scalar mass might be excluded by the current DD experiments. 
However, this conclusion no longer holds in the presence of additional dark matter components.
In our scenario, the two singlet fermions can account for a substantial fraction of the total relic abundance, thereby reducing the relic density fraction associated with the scalar component.
Consequently, $\xi_{\text{scalar}} \times \sigma^p_{\text{scalar}}$ becomes reasonably small for 
$m_\phi$ larger than the Higgs mass, pushing the cross section below the bound from XENONnT, yet 
above the neutrino floor \cite{Billard:2021uyg}.
Therefore, a viable scalar dark matter with mass range of approximately $125-400$ GeV 
remains below the current XENONnT limits while lying above the neutrino floor, making it 
potentially accessible to future direct detection experiments.
Since the region $m_\phi < m_h/2$ is already excluded by XENON1T constraints, we do not 
place additional bounds from invisible Higgs decay.

\section{Conclusion}
\label{Conc}
We have studied a minimal $\mathbb{Z}_2$-symmetric extension of the Standard Model containing two singlet Dirac fermions and a real singlet scalar.
The $\mathbb{Z}_2$ symmetry allows only a quadratic Higgs-portal interaction between the dark sector and the Standard Model. When appropriate kinematic conditions are imposed on the masses of the three dark-sector particles, the model realizes a three-component dark matter scenario.

We identified sizeable regions of parameter space consistent with the observed dark matter relic abundance measured by Planck. In these regions, the three dark matter components contribute differently to the total relic density, leading to a rich freeze-out phenomenology governed by the interplay of self-annihilation, conversion, and co-annihilation processes.

A distinctive feature of the model is that the scalar dark matter particle interacts with nucleons at tree level through the Higgs-portal, whereas the direct detection cross sections of the fermionic dark matter components arise only at one loop and are therefore strongly suppressed. Since the experimentally relevant direct detection rate is proportional to the relic density fraction of each component, viable regions of parameter space remain in agreement with current XENON1T and XENONnT constraints.

We find that the fermionic dark matter components can account for a substantial fraction of the observed relic abundance while possessing direct detection cross sections below the neutrino floor. At the same time, the scalar dark matter component can remain experimentally accessible. In particular, a viable scalar mass range of approximately $125-400$ GeV lies below the current direct detection limits while remaining above the neutrino floor, making it a promising target for future direct detection experiments.

\bibliography{ref}

@article{Alguero:2023zol,
    author = "Alguero, G. and Belanger, G. and Boudjema, F. and Chakraborti, S. and Goudelis, A. and Kraml, S. and Mjallal, A. and Pukhov, A.",
    title = "{micrOMEGAs 6.0: N-component dark matter}",
    eprint = "2312.14894",
    archivePrefix = "arXiv",
    primaryClass = "hep-ph",
    doi = "10.1016/j.cpc.2024.109133",
    journal = "Comput. Phys. Commun.",
    volume = "299",
    pages = "109133",
    year = "2024"
}

@article{Bergstrom:2000pn,
    author = {Bergstr{\"o}m, Lars},
    title = "{Nonbaryonic dark matter: Observational evidence and detection methods}",
    eprint = "hep-ph/0002126",
    archivePrefix = "arXiv",
    doi = "10.1088/0034-4885/63/5/2r3",
    journal = "Rept. Prog. Phys.",
    volume = "63",
    pages = "793",
    year = "2000"
}

@article{Feng:2010gw,
    author = "Feng, Jonathan L.",
    title = "{Dark Matter Candidates from Particle Physics and Methods of Detection}",
    eprint = "1003.0904",
    archivePrefix = "arXiv",
    primaryClass = "astro-ph.CO",
    reportNumber = "UCI-TR-2009-13",
    doi = "10.1146/annurev-astro-082708-101659",
    journal = "Ann. Rev. Astron. Astrophys.",
    volume = "48",
    pages = "495--545",
    year = "2010"
}

@article{Arcadi:2017kky,
    author = "Arcadi, Giorgio and Dutra, Ma{\'\i}ra and Ghosh, Pradipta and Lindner, Manfred and Mambrini, Yann and Pierre, Mathias and Profumo, Stefano and Queiroz, Farinaldo S.",
    title = "{The waning of the WIMP? A review of models, searches, and constraints}",
    eprint = "1703.07364",
    archivePrefix = "arXiv",
    primaryClass = "hep-ph",
    doi = "10.1140/epjc/s10052-018-5662-y",
    journal = "Eur. Phys. J. C",
    volume = "78",
    number = "3",
    pages = "203",
    year = "2018"
}

@article{Steigman:2012nb,
    author = "Steigman, Gary and Dasgupta, Basudeb and Beacom, John F.",
    title = "{Precise Relic WIMP Abundance and its Impact on Searches for Dark Matter Annihilation}",
    eprint = "1204.3622",
    archivePrefix = "arXiv",
    primaryClass = "hep-ph",
    doi = "10.1103/PhysRevD.86.023506",
    journal = "Phys. Rev. D",
    volume = "86",
    pages = "023506",
    year = "2012"
}

@article{Ghorbani:2018pjh,
    author = "Ghorbani, Karim and Ghorbani, Parsa Hossein",
    title = "{Leading Loop Effects in Pseudoscalar-Higgs Portal Dark Matter}",
    eprint = "1812.04092",
    archivePrefix = "arXiv",
    primaryClass = "hep-ph",
    doi = "10.1007/JHEP05(2019)096",
    journal = "JHEP",
    volume = "05",
    pages = "096",
    year = "2019"
}

@article{Ghorbani:2014qpa,
      author         = "Ghorbani, Karim",
      title          = "{Fermionic dark matter with pseudo-scalar Yukawa
                        interaction}",
      journal        = "JCAP",
      volume         = "1501",
      year           = "2015",
      pages          = "015",
      doi            = "10.1088/1475-7516/2015/01/015",
      eprint         = "1408.4929",
      archivePrefix  = "arXiv",
      primaryClass   = "hep-ph",
      SLACcitation   = "%%CITATION = ARXIV:1408.4929;%%"
}

@article{XENON:2018voc,
    author = "Aprile, E. and others",
    collaboration = "XENON",
    title = "{Dark Matter Search Results from a One Ton-Year Exposure of XENON1T}",
    eprint = "1805.12562",
    archivePrefix = "arXiv",
    primaryClass = "astro-ph.CO",
    doi = "10.1103/PhysRevLett.121.111302",
    journal = "Phys. Rev. Lett.",
    volume = "121",
    number = "11",
    pages = "111302",
    year = "2018"
}

@article{XENON:2023cxc,
    author = "Aprile, E. and others",
    collaboration = "XENON",
    title = "{First Dark Matter Search with Nuclear Recoils from the XENONnT Experiment}",
    eprint = "2303.14729",
    archivePrefix = "arXiv",
    primaryClass = "hep-ex",
    doi = "10.1103/PhysRevLett.131.041003",
    journal = "Phys. Rev. Lett.",
    volume = "131",
    number = "4",
    pages = "041003",
    year = "2023"  
}

@article{Billard:2021uyg,
    author = "Billard, Julien and others",
    title = "{Direct Detection of Dark Matter -- APPEC Committee Report}",
    eprint = "2104.07634",
    archivePrefix = "arXiv",
    primaryClass = "hep-ex",
    month = "4",
    year = "2021"  
}

@article{Fan:2015sza,
    author = "Fan, JiJi and Koushiappas, Savvas M. and Landsberg, Greg",
    title = "{Pseudoscalar Portal Dark Matter and New Signatures of Vector-like Fermions}",
    eprint = "1507.06993",
    archivePrefix = "arXiv",
    primaryClass = "hep-ph",
    doi = "10.1007/JHEP01(2016)111",
    journal = "JHEP",
    volume = "01",
    pages = "111",
    year = "2016"
}

@article{Yang:2016wrl,
    author = "Yang, Kwei-Chou",
    title = "{Fermionic Dark Matter through a Light Pseudoscalar Portal: Hints from the DAMA Results}",
    eprint = "1604.04979",
    archivePrefix = "arXiv",
    primaryClass = "hep-ph",
    reportNumber = "CYCU-HEP-16-03",
    doi = "10.1103/PhysRevD.94.035028",
    journal = "Phys. Rev. D",
    volume = "94",
    number = "3",
    pages = "035028",
    year = "2016"
}

@article{Baek:2017vzd,
    author = "Baek, Seungwon and Ko, P. and Li, Jinmian",
    title = "{Minimal renormalizable simplified dark matter model with a pseudoscalar mediator}",
    eprint = "1701.04131",
    archivePrefix = "arXiv",
    primaryClass = "hep-ph",
    doi = "10.1103/PhysRevD.95.075011",
    journal = "Phys. Rev. D",
    volume = "95",
    number = "7",
    pages = "075011",
    year = "2017"
}

@article{Ghorbani:2017jls,
    author = "Ghorbani, Parsa Hossein",
    title = "{Electroweak Baryogenesis and Dark Matter via a Pseudoscalar vs. Scalar}",
    eprint = "1703.06506",
    archivePrefix = "arXiv",
    primaryClass = "hep-ph",
    doi = "10.1007/JHEP08(2017)058",
    journal = "JHEP",
    volume = "08",
    pages = "058",
    year = "2017"
}

@article{DiazSaez:2021pmg,
    author = "D\'\i{}az S\'aez, Basti\'an and Escalona, Patricio and Norero, Sebasti\'an and Zerwekh, Alfonso R.",
    title = "{Fermion singlet dark matter in a pseudoscalar dark matter portal}",
    eprint = "2105.04255",
    archivePrefix = "arXiv",
    primaryClass = "hep-ph",
    doi = "10.1007/JHEP10(2021)233",
    journal = "JHEP",
    volume = "10",
    pages = "233",
    year = "2021"
}

@article{Chen:2024njd,
    author = "Chen, Yu-Tong and Matsumoto, Shigeki and Tang, Tian-Peng and Tsai, Yue-Lin Sming and Wu, Lei",
    title = "{Light thermal dark matter beyond p-wave annihilation in minimal Higgs portal model}",
    eprint = "2403.02721",
    archivePrefix = "arXiv",
    primaryClass = "hep-ph",
    doi = "10.1007/JHEP05(2024)281",
    journal = "JHEP",
    volume = "05",
    pages = "281",
    year = "2024"
}

@article{Kim:2008pp,
    author = "Kim, Yeong Gyun and Lee, Kang Young and Shin, Seodong",
    title = "{Singlet fermionic dark matter}",
    eprint = "0803.2932",
    archivePrefix = "arXiv",
    primaryClass = "hep-ph",
    doi = "10.1088/1126-6708/2008/05/100",
    journal = "JHEP",
    volume = "05",
    pages = "100",
    year = "2008"
}

@article{McDonald:1993ex,
    author = "McDonald, John",
    title = "{Gauge singlet scalars as cold dark matter}",
    eprint = "hep-ph/0702143",
    archivePrefix = "arXiv",
    reportNumber = "IFM-13-93",
    doi = "10.1103/PhysRevD.50.3637",
    journal = "Phys. Rev. D",
    volume = "50",
    pages = "3637--3649",
    year = "1994"
}

@article{Planck:2018vyg,
    author = "Aghanim, N. and others",
    collaboration = "Planck",
    title = "{Planck 2018 results. VI. Cosmological parameters}",
    eprint = "1807.06209",
    archivePrefix = "arXiv",
    primaryClass = "astro-ph.CO",
    doi = "10.1051/0004-6361/201833910",
    journal = "Astron. Astrophys.",
    volume = "641",
    pages = "A6",
    year = "2020",
    note = "[Erratum: Astron.Astrophys. 652, C4 (2021)]"
}

@article{Alguero:2022inz,
    author = "Alguero, Gael and Belanger, Genevieve and Kraml, Sabine and Pukhov, Alexander",
    title = "{Co-scattering in micrOMEGAs: A case study for the singlet-triplet dark matter model}",
    eprint = "2207.10536",
    archivePrefix = "arXiv",
    primaryClass = "hep-ph",
    doi = "10.21468/SciPostPhys.13.6.124",
    journal = "SciPost Phys.",
    volume = "13",
    pages = "124",
    year = "2022"
}

@article{Bhattacharya:2013hva,
    author = "Bhattacharya, Subhaditya and Drozd, Aleksandra and Grzadkowski, Bohdan and Wudka, Jose",
    title = "{Two-Component Dark Matter}",
    eprint = "1309.2986",
    archivePrefix = "arXiv",
    primaryClass = "hep-ph",
    reportNumber = "UCRHEP-T537",
    doi = "10.1007/JHEP10(2013)158",
    journal = "JHEP",
    volume = "10",
    pages = "158",
    year = "2013"
}

@article{Belanger:2014vza,
    author = "B{\'e}langer, G. and Boudjema, F. and Pukhov, A. and Semenov, A.",
    title = "{micrOMEGAs4.1: two dark matter candidates}",
    eprint = "1407.6129",
    archivePrefix = "arXiv",
    primaryClass = "hep-ph",
    doi = "10.1016/j.cpc.2015.03.003",
    journal = "Comput. Phys. Commun.",
    volume = "192",
    pages = "322--329",
    year = "2015"
}

@article{Collins:2016aya,
    author = "Collins, John C. and Vermaseren, J. A. M.",
    title = "{Axodraw Version 2}",
    eprint = "1606.01177",
    archivePrefix = "arXiv",
    primaryClass = "cs.OH",
    month = "5",
    year = "2016"
}
\bibliographystyle{utphys}

\end{document}